\documentclass[twocolumn,nofootinbib,amsmath,amssymb,a4paper]{revtex4}

\usepackage{graphicx}
\usepackage{dcolumn}
\usepackage{bm}

\begin{document}

\title{Distance preconditioning for lattice Dirac operators}

\author{G.M. de Divitiis$^{a,b}$, R. Petronzio$^{a,b}$, N. Tantalo$^{b,c}$}
\affiliation{\vskip 10pt
$^{a}$~Universit\`a di Roma ``Tor Vergata'', I-00133 Rome, Italy\\
$^{b}$~INFN sezione di Roma ``Tor Vergata'', I-00133 Rome, Italy\\
$^{c}$~Centro Enrico Fermi, I-00184 Rome, Italy
}%

\begin{abstract}
We propose a preconditioning of the Dirac operator based on the factorisation of a predefined function related to the decay of the propagator with the distance. We show that it can improve the accuracy of correlators involving heavy quarks at large distances and accelerate the computation of light quark propagators.
\end{abstract}

\maketitle

\section{introduction}
A key ingredient of lattice QCD simulations is the inversion of the Dirac operator
which enters the generation of unquenched gauge field configurations and the computation of hadronic observables. One needs to solve numerically a linear system of the form
\begin{eqnarray}
\sum_{y}(D[U]+M)_{x,y}\ S(y) = \eta(x)
\label{eq:linsysunprec}
\end{eqnarray}
where $D[U]$ is the chosen discretization of the massless interacting Dirac operator, $M$ is the quark mass in lattice units, $\eta(x)$ 
is a source vector that is different from zero on a single time-slice (that without
any loss we shall assume to be at $x_0=0$). 
The solution $S(y)$ is obtained by iterative numerical algorithms, solvers, devised to 
invert so-called sparse matrices, like the matrices that result from the discretization 
of differential equations by finite differences methods. In this letter we shall not discuss 
the details of any particular solver (see ref.~\cite{Luscher:2010ae} for a complete review 
and for an updated list of references). For any solver one checks if the condition
\begin{eqnarray}
\left\vert \sum_{y}(D[U]+M)_{x,y}\ S^n(y) - \eta(x) \right\vert < r
\label{eq:stopping}
\end{eqnarray}
is satisfyed. Here $S^n(y)$ is the tentative solution at iteration number $n$,
the norm is any good norm in field space and $r$, the residue, is the global numerical 
accuracy requested for the solution. Typically $r$ is a small number of the
order of the arithmetic precision allowed by the computer architecture.
 Depending on the values of the quark mass the solution of eq.~(\ref{eq:linsysunprec}) poses 
different numerical problems. For light quarks the matrix $(D[U]+M)_{x,y}$ is badly conditioned 
and its numerical inversion requires a big number of iterations. At the other extreme, the number of iterations required
for heavy quark masses is  small  but there may be problems with the
numerical accuracy resulting for the time-slices far away from the source ($y_0\gg 0$).
Indeed eq.~(\ref{eq:stopping}) is a \emph{global} condition while for heavy quark propagators
the time-slices far away from the source are exponentially suppresed by a factor of
the order of $\exp(-My_0)$ and give a negligible contribution to the norm on the left 
side of eq.~(\ref{eq:stopping}).
When this problem arises one cannot trust numerical results at large times and it becomes 
impossible to extract physical informations by fitting the leading exponentials contributing 
to correlation functions.

In order to alleviate both difficulties, we propose a preconditioning of the Dirac operator 
that factorises from the propagator a function aiming to modify its leading decay 
with the distance\footnote{see refs.~\cite{Juttner:2005ks,Joo:2009hq} for different approaches}.
The simplest choice is to factorize a function $\alpha(y_0)$, to  solve numerically 
the preconditioned equation, and to restore the original propagator by multiplying each 
time slice for $1/\alpha(y_0)$.  $\alpha(y_0)$ is defined such that all the
different time-slices give comparable contributions to the calculation of the residue in 
the preconditioned case.
Our preconditioning is inspired to what is usually done in deriving the Eichten and 
Hill~\cite{Eichten:1989zv} lattice HQET action but of course does not introduce any approximation. 
Indeed, the choice above is suited for heavy quark propagators, while for light quark masses we will introduce a generalisation of the factorised function.

\begin{figure*}
\begin{center}
\includegraphics[width=0.9\textwidth]{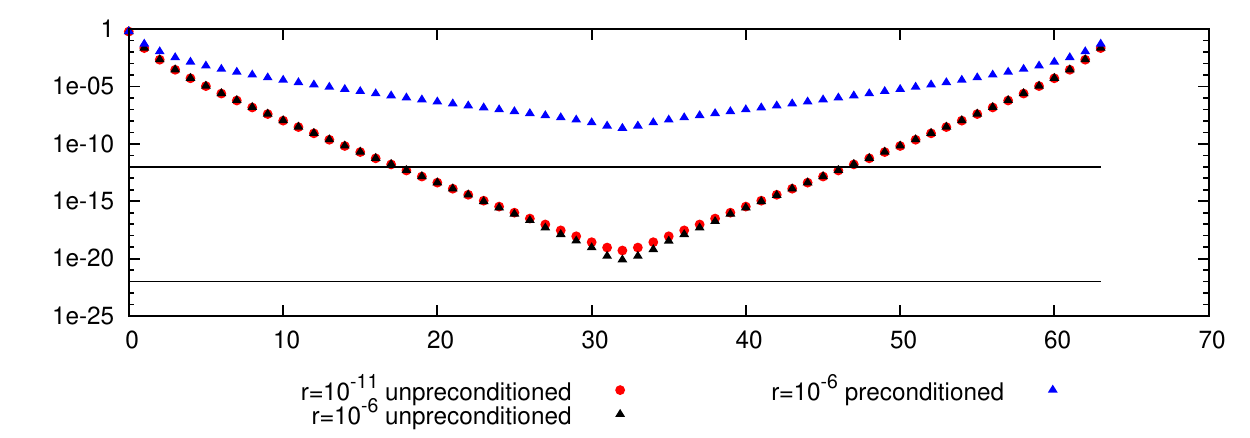}
\caption{\label{fig:treeuncorrect1} 
The red points correspond to the correlation fuction
$-C_{PP}(y_0)$ obtained by inverting the lattice Dirac equation
in the free theory with $M\simeq 0.5$ and with a residue $r=10^{-11}$.
The black points correspond to the same quantity but have been obtained
with a residue $r=10^{-6}$. The blue points correspond to the correlation
function $-C_{PP}^\prime(y_0)$ obtained by solving the preconditioned 
lattice Dirac equation with $M\simeq 0.5$ and $\alpha_0=0.4$. The two
black lines correspond to $r^2$ for the two values of the residue used in
the calculations. We use logarithmic scale on the $y$-axis.
}
\end{center}
\end{figure*}
\begin{figure*}
\begin{center}
\includegraphics[width=0.9\textwidth]{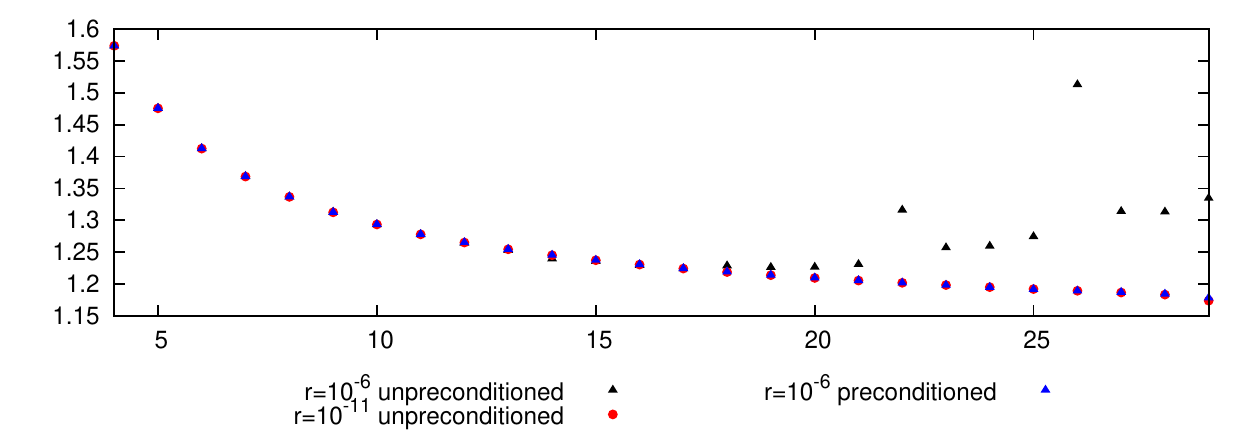}
\caption{\label{fig:treeuncorrect2} 
The red and black sets of points are the effective masses
of the corresponding correlators shown in FIG.~\ref{fig:treeuncorrect}. 
The blue set of points is the effective mass of the corresponding 
correlator in the top panel multiplyied
for the restoration factor as explained in the text.
}
\end{center}
\end{figure*}

\section{preconditioning heavy quark propagators}\label{sec:hproblem}

We work with the $O(a)$-improved Wilson lattice Dirac operator but the 
numerical problems that we address arise also with alternative discretisations of
the continuum action and the proposed solution can as well be easily implemented 
in those cases.
We have tested our preconditioning scheme both for heavy and for light quark masses
and in the free and in the interacting case. 
We start with the results for the heavy quarks. 
We first want to pick up a case where the problem arises. 
As an example, we have calculated the correlation function
\begin{eqnarray}
C_{PP}(y_0)=-\sum_{\vec{y}}{\mbox{tr}\left[S^{\dagger}(y)S(y)\right]}
\end{eqnarray}
by solving eq.~(\ref{eq:linsysunprec}) for a heavy quark propagator
of mass $M\simeq 0.5$ in the free theory  
for different choices of the residue. More precisely the red points in FIG.~\ref{fig:treeuncorrect1} have been obtained with a residue $r=10^{-11}$ while the black points with a residue $r=10^{-6}$
and the two black lines correspond to the squares of these two values of $r$. 
As is clearly visible from FIG.~\ref{fig:treeuncorrect1},
and from FIG.~\ref{fig:treeuncorrect2} where we show the effective masses of the correlations
shown in FIG.~\ref{fig:treeuncorrect1}, the black points start to deviate from the red ones for 
$y_0\simeq 18$, i.e. when the correlator, which in this case is just the square module of the propagator,  becomes smaller than the square of the "loose" residue $r=10^{-6}$. 

If the time extent of the lattice is not too large the problem can be solved by brute force by 
lowering the residue and the results obtained in the preset case with $r=10^{-11}$ can
be considered as exact. If instead the time extent of the lattice is rather large the brute force approach cannot be considered because the required
residues would be smaller than what is allowed on double-precision architectures, even
in the case of moderately heavy quarks. In the case under consideration, by choosing a loose precision, i.e. a residue $r=10^{-6}$, we make the numerical problem evident
and we show that also such an "extreme" situation can be recovered by using our proposal.
Moreover, we notice that a residue $r=10^{-6}$ is the smallest   allowed on single-precision architectures that
presently are considerably much faster than double-precision ones.

We now come to the proposed solution. We redefine the quark fields and the propagators
as follows
\begin{eqnarray}
S(y)&=&\alpha(y_0)\ S^\prime(y)
\nonumber \\
\nonumber \\
\eta(y)&=&\alpha(y_0)\ \eta^\prime(y)
\label{eq:redeft}
\end{eqnarray}
Once the previous expressions are inserted in eq.~(\ref{eq:linsysunprec}) we get
the preconditioned system
\begin{eqnarray}
\sum_{y}(D^\prime[U]+M)_{x,y}\ S^\prime(y) = \eta^\prime(x)
\label{eq:linsysprec}
\end{eqnarray}
that we solve numerically in place of eq.~(\ref{eq:linsysunprec}). In order to
write the preconditioned Dirac operator it is sufficient to modify the
forward and backward lattice covariant derivatives in the time direction
accoring to
\begin{eqnarray}
\nabla_0S(y) &=& U_0(y)S(y+\hat{0})-S(y) 
\nonumber \\
\nonumber \\
&\rightarrow&
\frac{\alpha(y_0+1)}{\alpha(y_0)}U_0(y)S^\prime(y+\hat{0})-S^\prime(y)
\nonumber \\
\nonumber \\
\nonumber \\
\nonumber \\
\nabla_0^{\dagger}S(y) &=& S(y)-U_0^{\dagger}(y-\hat{0})S(y-\hat{0}) 
\nonumber \\
\nonumber \\
&\rightarrow&
S^\prime(y)-\frac{\alpha(y_0-1)}{\alpha(y_0)}U_0^{\dagger}(y-\hat{0})S^\prime(y-\hat{0})
\nonumber \\
\nonumber \\
\end{eqnarray}
Particular care has to be used at the boundaries of the lattice in order
to respect the boundary conditions originally satisfied by the quark fields.
If as in the case of FIG.~\ref{fig:treeuncorrect1} $S(y)$ satisfies anti-periodic boundary 
conditions along the time direction, it follows from eq.~(\ref{eq:redeft}) that
\begin{eqnarray}
&&S(y+L_0\hat{0})=-S(y)
\nonumber \\
\nonumber \\
&&S^\prime(y+L_0\hat{0})=-\frac{\alpha(y_0)}{\alpha(y_0+L_0\hat{0})}S^\prime(y)
\label{eq:redeftbc}
\end{eqnarray}

The blue points in FIG.~\ref{fig:treeuncorrect1} correspond to the correlation function
\begin{eqnarray}
C_{PP}^\prime(y_0)=-\sum_{\vec{y}}{\mbox{tr}\left[(S^\prime)^{\dagger}(y)S^\prime(y)\right]}
\end{eqnarray}
obtained by solving eq.~(\ref{eq:linsysprec}) with the loose residue $r=10^{-6}$ but
after having factorized the function
\begin{eqnarray}
\alpha(y_0)=\cosh[\alpha_0(y_0-L_0/2)]
\label{eq:pbcalpha}
\end{eqnarray}
by setting $\alpha_0=0.4$. As expected, the preconditioned correlator stays above the 
line of the loose precision residue and the "exact" result can be back recovered as follows
\begin{eqnarray}
C_{PP}(y_0)=\left[\alpha(y_0)\right]^2\ C_{PP}^\prime(y_0)
\label{eq:backcorrection}
\end{eqnarray}
In
FIG.~\ref{fig:treeuncorrect2} the blue points correspond to
the effective mass of the preconditioned correlator after the "restoration"
of eq.~(\ref{eq:backcorrection}) and fall exactly on top of
the red ones in spite of the fact that they have been obtained with the same loose precision that affected the non preconditioned black points.

\begin{figure*}
\begin{center}
\includegraphics[width=0.9\textwidth]{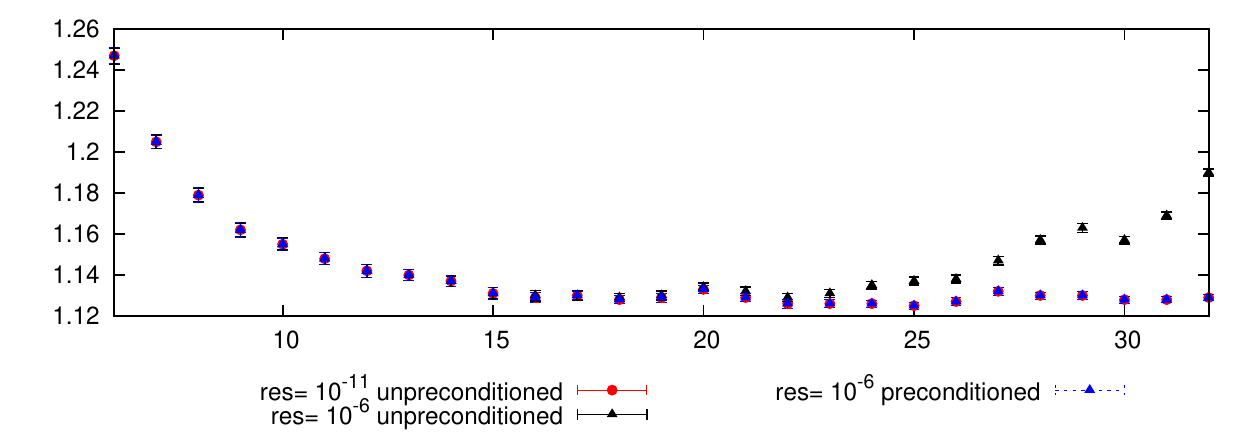}
\caption{\label{fig:iuncorrect} 
The red points correspond to the effective mass of the correlation fuction
$-C_{PP}(y_0)$ obtained by inverting the lattice Dirac equation
in the interacting theory with $am^{pcac}_h\simeq 0.35$ and with a residue $r=10^{-11}$.
The black points correspond to the same quantity but have been obtained
with a residue $r=10^{-6}$. The blue points correspond to the effective mass
of the restored correlation function $-C_{PP}^\prime(y_0)$ obtained by solving 
the preconditioned lattice Dirac equation with $am^{pcac}_{h}\simeq 0.35$ and $\alpha_0=0.4$.
}
\end{center}
\end{figure*}

In FIG.~\ref{fig:iuncorrect} we show the same plot as in 
FIG.~\ref{fig:treeuncorrect2} but in the interacting theory. The gauge ensamble
used correspond to the entry $E5$ in TABLE~\ref{tab:iterations}.
The size of the lattice is $L_0L_1L_2L_3=64\times 32^3$ and the hopping parameter
of the sea quarks is $k_{sea}=0.13625$ corresponding to a PCAC quark mass of about
$am^{PCAC}_{sea}\simeq 0.07$. The data shown in FIG.~\ref{fig:iuncorrect} correspond
to a pseudoscalar-pseudoscalar correlator, as in the free theory case, of
two degenerate heavy quarks with hopping parameters $k_h=0.125$ corresponding
to a PCAC quark mass of about $am^{PCAC}_{h}\simeq 0.35$. The unpreconditioned
correlators decay approximately as fast as in the free theory case and from the
difference of the black (unpreconditioned, $r=10^{-6}$) and red (unpreconditioned, $r=10^{-11}$) 
sets of data we see the same distortion of FIG.~\ref{fig:treeuncorrect2}.
The blue points have been obtained by solving eq.~(\ref{eq:linsysprec}) after having
factorized $\alpha(y_0)$ with $\alpha_0=0.4$ and by restoring the results
according to eq.~(\ref{eq:backcorrection}). Also in the interacting theory the
blue points are identical to red points though they have been obtained with
the same loose residue $r=10^{-6}$ used to obtain the black points.

We close this section by observing that our preconditioning technique may
be particularly useful when working with the Schr\"odinger Functional~\cite{Luscher:1992an,Sint:1993un} 
formulation of the theory. In this case, countrary to the case of periodic boundary conditions
along the time direction,  the correlators decay exponantially over the whole
time extent of the lattice and one has to choose very small residues also in computing
relatively light quark propagators. We have performed several succesful experiments with our 
preconditioning technique also in the Schr\"odinger Functional case by using
$\alpha(x_0)=\exp(-\alpha_0 x_0)$.

\section{preconditioning light quark propagators}\label{sec:lproblem}
  \begin{table}[t]
  \begin{center}
  \begin{tabular}{ccccccc}
  $\qquad\qquad$& $\beta$ & $L_0L_1L_2L_3$ & $k_{sea}$  & $\qquad r \qquad$ & 
  $\alpha_0$ & iterations\\
  \hline
  \\
  $D5$ & 5.3 & $48\times 24^3$   & 0.13625 &  $10^{-11}$ & 0.0 & 175 \\
  $D5$ & 5.3 & $48\times 24^3$   & 0.13625 &  $10^{-11}$ & 0.4 & 141 \\[10pt]

  $E3$ & 5.3 & $64\times 32^3$   & 0.13605 &  $10^{-10}$ & 0.0 &  99\\
  $E3$ & 5.3 & $64\times 32^3$   & 0.13605 &  $10^{-10}$ & 0.2 &  78\\
  $E3$ & 5.3 & $64\times 32^3$   & 0.13605 &  $10^{-10}$ & 0.4 &  69\\[10pt]

  $E4$ & 5.3 & $64\times 32^3$   & 0.13610 &  $10^{-10}$ & 0.0 & 115\\
  $E4$ & 5.3 & $64\times 32^3$   & 0.13610 &  $10^{-10}$ & 0.2 &  91\\
  $E4$ & 5.3 & $64\times 32^3$   & 0.13610 &  $10^{-10}$ & 0.4 &  81\\[10pt]

  $E5$ & 5.3 & $64\times 32^3$   & 0.13625 &  $10^{-10}$ & 0.0 & 194\\
  $E5$ & 5.3 & $64\times 32^3$   & 0.13625 &  $10^{-10}$ & 0.2 & 153\\
  $E5$ & 5.3 & $64\times 32^3$   & 0.13625 &  $10^{-10}$ & 0.4 & 141\\[10pt]
  \hline
  \end{tabular}
  \end{center}
  \caption{\label{tab:iterations}
  Gauge configurations have been generated with $n_f=2$ dynamical
  $O(a)$-improved Wilson quarks with $c_{sw}=1.90952$. The figures in
  the last column correspond to the average of the number of iterations 
  required to invert the Dirac equation in the unitary theory by using
  the SAP+GCR inverter for several values of the preconditioning parameter
  $\alpha_0$. The values corresponding to $\alpha_0=0.0$ have been obtained
  without using our preconditioning technique.
  }
  \end{table}   

In this section we shall briefly discuss how the ideas developed and discussed
in the previous section can be used to accelerate the numerical calculation
of light quark propagators. We start our discussion by generalizing eq.~(\ref{eq:redeft})
as follows
\begin{eqnarray}
S(y)&=&\beta(y_0,y_1,y_2,y_3)\ S^\prime(y)
\nonumber \\
\nonumber \\
\eta(y)&=&\beta(y_0,y_1,y_2,y_3)\ \eta^\prime(y)
\label{eq:redefa}
\end{eqnarray}
In the following we shall consider the particular choice
\begin{eqnarray}
\beta(y_0,y_1,y_2,y_3) &=& \prod_{\mu=0}^3{\alpha(y_\mu)}
\nonumber \\
\nonumber \\
&=& \prod_{\mu=0}^3{\frac{1}{\cosh[\alpha_0(y_\mu-L_\mu/2)]}}
\label{eq:betaa}
\end{eqnarray}
and the preconditioned lattice Dirac operator can be obtained as easily as before
by changing all the covariant derivatives according to
\begin{eqnarray}
\nabla_\mu S(y) &=& U_\mu(y)S(y+\hat{\mu})-S(y) 
\nonumber \\
\nonumber \\
&\rightarrow&
\frac{\alpha(y_\mu+1)}{\alpha(y_\mu)}U_\mu(y)S^\prime(y+\hat{\mu})-S^\prime(y)
\nonumber \\
\nonumber \\
\nonumber \\
\nonumber \\
\nabla_\mu^{\dagger}S(y) &=& S(y)-U_\mu^{\dagger}(y-\hat{\mu})S(y-\hat{\mu}) 
\nonumber \\
\nonumber \\
&\rightarrow&
S^\prime(y)-\frac{\alpha(y_\mu-1)}{\alpha(y_\mu)}U_\mu^{\dagger}(y-\hat{\mu})S^\prime(y-\hat{\mu})
\nonumber \\
\nonumber \\
\label{eq:redefcovs}
\end{eqnarray}
and by changing accordingly the boundary conditions in all directions as
done in eqs.~(\ref{eq:redeftbc}) for the time direction. 

An important difference of the present case with respect to the one discussed in
the previous section is that the restoration of the true propagator must be performed
 before making the contractions needed to build correlation
functions by using the first of eqs.~(\ref{eq:redefa}). 

Here the preconditioning is not to gain precision, but to accelerate the convergence of the inversion. 
Therefore, by applying  eqs.~(\ref{eq:redefa}), (\ref{eq:betaa}) and
(\ref{eq:redefcovs}) to the calculation of a light quark propagator one aims to
make the propagator to decay faster than the original
unpreconditioned operator. By judiciously chosing the parameter $\alpha_0$ 
it is possible to change the condition number of the preconditioned
system without compromising the numerical accuracy of the solution,
an operation that should be performed on double-precision computer architectures.

In TABLE~\ref{tab:iterations} we quantify the gain in computational time that can
be achieved by showing the number of iterations of the SAP+GCR solver required to
solve the lattice Dirac equation for light quarks with and without 
our preconditioning. The SAP+GCR solver has been introduced and explained
in details by the author in ref.~\cite{Luscher:2003qa}. The 
gauge ensambles used to perform this test have been
generated within the CLS agreement~\cite{CLS} with the parameters given in the table. In the case of the $E$-lattices 
the SAP+GCR solver has been ran on $128$ processors of a cluster of PC's by dividing 
the global lattices into blocks of $4^4$ points. In the case of the $D$-lattice the 
SAP+GCR solver has been ran on $32$ processors of a cluster of PC's by dividing the global
lattices into blocks of $6^4$ points.
The table shows that by increasing
the value of the parameter $\alpha_0$ the number of iterations goes down with
a time gain that can easily reach the $30\%$. In the case under discussion, we checked that higher values of $\alpha_0$ would induce a "heavy quark" like behavior and produce distorted results for the reasons discussed at lenghty in the previous section. 

The method discussed in this letter can be generalised by adding some Dirac structure in the 
factorised function, an option presently under investigation.

\begin{acknowledgments}
We thanks our colleagues of the CLS community for sharing the efforts required to generate
the dynamical gauge ensambles used in this study.
\end{acknowledgments}


\end{document}